\documentclass[ aps,%
floatfix,%
final,%
notitlepage,%
oneside,%
onecolumn,%
nobibnotes,%
nofootinbib,%
superscriptaddress,%
showpacs,%
]%
{revtex4}

\usepackage{epsfig}
\usepackage{graphics}

\newcommand{\JP}{\psi} \newcommand{\ec}{\eta_c}

\newcommand{\epem}{e^+e^-}

\newcommand{\fb}{\,\mathrm{fb}}

\newcommand{\beq}{\begin{eqnarray}}\newcommand{\eeq}{\end{eqnarray}}
\newcommand{\beqa}{\begin{eqnarray*}}\newcommand{\eeqa}{\end{eqnarray*}}

\begin{document}

\title{Excited charmonium mesons production in $e^+e^-$ annihilation at $\sqrt{s}=10.6$ GeV}
\author{V.V. Braguta}
\email{braguta@mail.ru}

\author{A.K. Likhoded}
\email{Likhoded@ihep.ru}

\author{A.V. Luchinsky}
\email{Alexey.Luchinsky@ihep.ru}
\affiliation{Institute for High Energy Physics, Protvino, Russia}

\begin{abstract}
In this paper the production of excited vector and pseudoscalar charmonium mesons in $\epem$ annihilation is analyzed in the framework of light cone. 
In particular the cross sections $\epem \to \psi(2S)\eta_c(1S), \psi(1S)\eta_c(2S),\psi(2S)\eta_c(2S)$ have been calculated. It is shown, that contrary to NRQCD the cross sections calculated in the framework of light cone agree with experimental data. 
\end{abstract}

\pacs{
12.38.-t,  
12.38.Bx,  
13.66.Bc,  
13.25.Gv 
}

\maketitle

\newcommand{\ins}[1]{\underline{#1}}
\newcommand{\subs}[2]{\underline{#2}}
\section{Introduction}

$J/\psi, \ec$ mesons production  in $e^+ e^-$ annihilation at energy $\sqrt s = 10.6$ GeV remains 
very interesting task for theoretical investigations. The cross section of the process $\epem \to \JP \ec$ 
measured at Belle experiment was first presented in paper \cite{Abe:2002rb}. The lower bound of the cross section
measured at Belle
\beqa
  \sigma(\epem\to\JP\ec) & > & 33\fb
\eeqa
is about an order of magnitude higher, than the theoretical predictions \cite{Braaten:2002fi} obtained in the 
framework of NRQCD\cite{Bodwin:1994jh}. Some efforts were made to explain this discrepancy. 
For example, in \cite{Bodwin:2002fk} it was assumed that some of Belle's $J/\psi\eta_c$ signal could 
actually be $J/\psi J/\psi$ events but later in \cite{Bodwin:2002kk,Luchinsky:2003yh} it was shown that 
QCD corrections decrease the value of $\epem\to J/\psi J/\psi $ cross section and subsequent Belle analysis
\cite{Abe:2003ja} excluded this possibility completely. Among the other possible explanations the contributions 
from glueball \cite{Lee:2003db} or color-octet states. See also a complete review of the Quarkonium Working Group \cite{Brambilla:2004wf} and refrences therein.

There was a hope for the improvement of theoretical prediction by higher order QCD corrections. As was shown in paper 
\cite{Zhang:2005ch} QCD corrections really increase the cross section by a factor 1.8, but this is still insufficient to 
reach experimental results obtained at Belle.

Recently a surprisingly simple solution of this problem was found. It turns out that taking into account the 
intrinsic motion of quarks inside charmonium mesons one can significantly increase the value of cross section. 
First this effect was observed in the framework of light cone expansion method in works \cite{Ma:2004qf,Bondar:2004sv}. 
In paper  \cite{Braguta:2005gw} this effect was considered as the expansion of the amplitude in relative velocity of quark
inside mesons.

In the last paper it was proved that NRQCD series for the amplitude of the process $\epem\to\JP\ec$ in relative velocity of quark-antiquark pair 
in the $\JP, \ec$ mesons converges slowly (in potential models relative velocity for these mesons is $v \sim 0.5$ \cite{Buchmuller:1980su}). 
The reason of such behavior consists in strong dependence of quark and gluon 
propagators in the diagrams of the process from  relative momentum of quark-antiquark pair in meson. 
In order to show that this effect really takes place for the process $\epem\to\JP\ec$
the amplitude was expanded in relative velocity series except the propagators 
of intermediate particles for which the exact expression was used. As the result 
NRQCD prediction is multiplied by the factor that regards internal motion of quark-antiquark pair 
inside mesons and the cross section becomes in 2-5 times greater than that in the framework of NRQCD depending 
on the width of the wave function used. Thus one can conclude that the usage of
the leading approximation of NRQCD proved to be unreliable for $\epem \to \JP \ec$ at energy $\sqrt s = 10.6$GeV. 

As it was mentioned already the other method that can be used for theoretical prediction of  cross section of 
the process $\epem\to\JP\ec$ is light cone expansion method. The cross section calculated in the framework 
of light cone does not contradict to the experiment data. Unfortunately in papers \cite{Ma:2004qf,Bondar:2004sv} this method 
was used only for the calculation of the $\epem\to\JP\ec$ cross section. It would be interesting to look how 
this method works in other reactions measured at the experiments.  For instance in addition to the process $\epem\to\JP\ec$ Belle 
collaboration has measured the cross sections of the processes 
$\epem\to \JP(2S) \ec, \JP \ec(2S), \JP(2S) \ec(2S), \JP \chi_{c0}, \JP(2S) \chi_{c0}$\cite{Abe:2004ww}. 
Lately BaBar experiment has measured the cross sections of the processes 
$\epem\to \JP \ec, \JP \ec(2S), \JP \chi_{c0}$\cite{Aubert:2005tj}. 
In the frame work of NRQCD these processes were considered in paper\cite{Braaten:2002fi}. As 
in the case of $\epem\to\JP\ec$ Belle and BaBar results are in contradiction with 
NRQCD predictions. In our paper we will consider leading order contribution to the processes $\epem\to \JP(2S) \ec, \JP \ec(2S), \JP(2S) \ec(2S)$ 
in the framework of light cone. As to the processes $\epem\to  \JP \chi_{c0}, \JP(2S) \chi_{c0}$
we will argue that the leading order contribution to the cross section in $1/s$ series is much less 
than NLO. We are going to calculate NLO contribution in our forthcoming publication. 

This paper is organized as follows. Next section is devoted to the consideration 
of the processes $\epem\to \JP(2S) \ec, \JP \ec(2S), \JP(2S) \ec(2S)$ in the framework 
of light cone. In section 3 we argue that light cone LO contribution to the cross sections $\epem\to  \JP \chi_{c0}, \JP(2S) \chi_{c0}$
is much less than NLO one. In last section we summarize our results.

\section{The processes $\epem\to\JP\ec, \JP(2S) \ec, \JP \ec(2S), \JP(2S) \ec(2S)$. }

In this section the processes $\epem\to\ V P$, where $V=\JP(1S), \JP(2S)$, $P=\ec(1S), \ec(2S)$ will be considered.
The cross section of the process can be written as follows
\beq
\sigma (\epem\to\ V P) & = & \frac {\pi \alpha^2 q_c^2} {6}  \biggl ( \frac {2 |\mathbf{p}|} {\sqrt s}  \biggr )^3 |F_{vp}|^2,
\eeq
where $\sqrt s$ is invariant mass of $e^+ e^-$ system, $p$ ~is the momentum of final mesons in the center mass
frame. The formfactor $F_{vp}$ is defined in the following way 
\beq
\left<V(p_1, \lambda), P(p_2)| J_{\mu} | 0\right> & = & 
  \epsilon_{\mu \nu \rho \sigma} e^{\nu} p_1^{\rho} p_2^{\sigma} F_{vp}.
\label{Fvp}
\eeq
The asymptotic behavior of formfactor $F_{vp}$ with mesons in the final state can be obtained from the following formula\cite{Chernyak:1977as}
\beq
\left<M(p_1, \lambda_1) M(p_2, \lambda_2)|J_{\mu}|0\right> & \sim & 
  \left( \frac 1 {\sqrt s}  \right)^{|\lambda_1 - \lambda_2|+1}.
\label{as}
\eeq
Obviously in the case $M(p_1, \lambda_1)=V(p_1, \lambda), M(p_2, \lambda_2)=P(p_2)$ the helicity $\lambda_2=0$. As to 
the helicity $\lambda_1$ it is seen from formula (\ref{Fvp}) that vector meson $V$ is transversely polarized $\lambda_1= \pm 1$.
So the asymptotic behavior of the amplitude is
\beq
\left<V(p_1, \lambda), P(p_2)| J_{\mu} | 0\right> & \sim & 1/s,
\eeq
or $F_{vp} \sim 1/s^2$ is the asymptotic behavior of the formfactor.

Two diagrams that give contribution to the amplitude of the processes under consideration are 
presented in Fig 1. The other two diagrams can be obtained from the depicted ones by charge conjugation. 
The leading order contribution to the formfactor was first obtained in papers\cite{Ma:2004qf,Bondar:2004sv}  where the process
$\epem\to\JP\ec$ was considered. In our paper we follow work \cite{Bondar:2004sv}. In deriving the expression for the formfactor $F_{vp}$ 
in paper \cite{Bondar:2004sv} the mass difference of the final mesons
$\JP$ and $\ec$ was disregarded. The mass difference of  $\JP$ and $\ec$ mesons is about $\sim 100$MeV
and this value can not give large correction to the cross section. But if, for instance, $\JP(2S)$ and $\ec$ mesons 
are considered the mass difference is about $\sim 700$MeV. As it will be seen from the subsequent analysis this value is large enough to give large correction to the cross section under 
consideration. So we have derived the formula for the formfactor taking into the account different masses of final mesons. 
The expression for the formfactor $F_{vp}$ can be written as follows
\beq
|F_{vp}(s)| & = & \frac{32\pi}{9}   \left|\frac{f_V f_P  M_P M_V}{q_0^4}\right|\,I_0\,,
\label{res}
\eeq
\beqa
I_0 & = &
  \int^1_0 dx_1 \int^1_0 dy_1 \alpha_s(k^2) \left\{
    \frac {M_P} {M_V^2} \frac{Z_t Z_p 
    V_{T}(x) P_{P}(y)}{d(x,y)\, s(x)}- \frac 1 {M_P} \frac{\overline {M}_Q^2 }{{ M_V}^2}\,
    \frac{Z_m^\sigma Z_t V_T(x) P_A(y)}{d(x,y)\,s(x)}+
\right.\\ & + &
\frac{1}{2 M_P}\frac{V_{L}(x)\,P_{A}(y)}{d(x,y)}+\frac{1}{2 M_P}
\frac{(1-2y_1)}{s(y)}\frac{V_{\perp}(x)\,P_{A}(y)}{d(x,y)}+
\\ & + & \left.
\frac{1}{8} \biggl ( 1-Z_t Z_m^k\frac{4{ \overline M}_Q^2}{{ M_V}^2 }\biggr ) \frac 1 M_P \, 
\frac{(1+y_1)V_A(x)P_A(y)}
{d^2(x,y)}\right\}.
\eeqa
Where $q_0^2 \simeq (s-M_V^2-M_P^2)$, $P_A, P_P, V_T, V_L, V_{\perp}, V_A$ are 
the light cone wave functions defined in \cite{Bondar:2004sv}, $M_V, M_P$ are the mass of the 
vector and pseudoscalar mesons correspondingly, $\overline {M}_Q= M^{\overline {MS} }_Q ( \mu = M^{\overline {MS}}_Q )$, $Z_{t}$ and $Z_{p}$ are the
renormalization factors of the local tensor and pseudoscalar currents, $d(x,y), s(x), s(y)$ are defined as follows:
\beqa
d(x,y)& = & 
  \frac{k^2}{q_0^2}=\left( x_1+\frac{\delta}{y_1}\right)
  \left(y_1+\frac{\delta}{x_1}\right),
\qquad \delta=\Biggl (Z_m^k \frac{{\overline  M}_Q}{q_0}\Biggr )^2\,, 
\\ 
s(x) & = & \left(x_1+\frac{(Z_m^{\sigma}{\overline  M}_Q)^2}{y_1y_2\,q_o^2} \right),
\quad s(y)= \left(y_1+\frac{(Z_m^{\sigma}{\overline  M}_Q)^2}{x_1x_2\,q_o^2}
\right),
\\
Z_p & = & \left[\frac{\alpha_s(k^2)}{\alpha_s({\overline M}_Q^2)} \right]^{\frac{-3C_F}{b_o}},\quad
 Z_{t}=\left[\frac{\alpha_s(k^2)}{\alpha_s({\overline M}_Q^2)} \right]^{\frac{C_F}{b_o}},\quad
 Z_{m}(\mu^2)=\left[\frac{\alpha_s(\mu^2)}{\alpha_s({\overline M}_Q^2)} \right]^{\frac{3C_F}{b_o}},
\\ 
M_Q(\mu^2)& = & Z_m(\mu^2) M_Q,\quad \quad Z_m^k=Z_m(k^2),\,
\quad Z_m^{\sigma}=Z_m(\sigma^2)\,,
\eeqa
where ${ M}_Q(\mu^2)$ is the running $ {MS}$-mass, $C_F=4/3,\,b_o=25/3$,  $k=(k_1+l_1)$ is virtual gluon momentum, 
$\sigma=-k_1-l_1-l_2$ is virtual quark momentum in Fig. 1b.
For the light cone wave functions  $P_A, P_P, V_T, V_L, V_{\perp}, V_A$ of  $1S$ state mesons the following expressions will be used:
\beq
\phi_{i}(x,v^2) &=& 
  c_{i}(v^2)\,\phi^{a}_{i}(x)\left\{
    \frac{x_1 x_2}{[1-4x_1x_2(1-v^2)]} 
  \right\}^{1-v^2}\,
  \label{func1}
\eeq
where $v$ is a characteristic speed of quark-antiquark pair in meson, $c_{i}$ is 
the coefficient which is fixed by the wave function normalization $\int d x \phi_{i}(x,v^2) =1$, $\phi^{a}$
is the asymptotic expression for the wave function

In order get the wave functions of $2S$ states the following 
procedure will be used. We recall that $2S$ state Coulomb wave function has the form:
\begin{eqnarray}
\Psi_{2 S} (r) \sim (1- q_0 r) \exp(-q_0 r)= \biggl (1+ q_0 \frac d  {d q_0} \biggr ) \Psi_{1S} (r),
\end{eqnarray}
where $q_0= q_B/2$ is the mean momentum of quark inside meson, $\Psi_{1S} (r)$ is the $1S$ Coulomb wave function 
 with Born momentum equals $q_0$. In momentum space $2S$ wave function has the form:
\beq
\phi ( p) \sim \biggl (1 + q_0 \frac d  {d q_0} \biggr ) \frac 1 { (p^2 + q_0^2)^2 } = \frac {p^2 - 3 q_0^2} {(p^2+ q_0^2)^2}
\label{ff}
\eeq
$2S$ wave function (\ref{ff}) has zero at $p^2 = 3 q_0^2$. Obviously this zero has nothing 
to do with real zero of $c \bar c( 2S)$ meson wave function. In order to connect wave function (\ref{ff}) 
with more realistic model we replace $1 + q_0 \frac d  {d q_0} $ by $1 + \beta q_0 \frac d  {d q_0} $.
The constant $\beta$ is fixed by the condition that zero of the modified wave function  must 
coincide with zero obtained from the solution of Schrodinger equation with potential \cite{Buchmuller:1980su}.
Thus we obtain $\beta=0.38$. Now it is easy to find wave function of $2S$ state:
\beq
\phi ( p) \sim \biggl (1 + \beta q_0 \frac d  {d q_0} \biggr ) \Biggl \{ \frac{x_1 x_2} {[1-4x_1x_2(1-v^2)]} \Biggr \}^{\alpha}=
\biggl ( 1 - 8  v^2 \beta \frac { \alpha x_1 x_2} {[1-4x_1x_2(1-v^2)]} \biggr) \Biggl \{ \frac{x_1 x_2} {[1-4x_1x_2(1-v^2)]} \Biggr \}^{\alpha}
\eeq
The constant $\alpha$ equals unity in the case of usual Coulomb wave function. In paper \cite{Bondar:2004sv} the constant $\alpha$
was taken $1-v^2$ (\ref{func1}) since this value allows one to link different behavior of the wave function: $v \to 0$ and $v \to 1$.
In our analysis we will take the same value of this constant. Finally one gets
 the light cone wave functions $P_A, P_P, V_T, V_L, V_{\perp}, V_A$ of $2S$ state mesons  
\beq
\phi_{i}(x,v^2)=c_{i}(v^2)\,\phi^{a}_{i}(x)\, \Biggl ( 1 - 8  v^2 \beta \frac {(1-v^2) x_1 x_2} {[1-4x_1x_2(1-v^2)]} \Biggr) \Biggl \{ \frac{x_1 x_2}
{[1-4x_1x_2(1-v^2)]} \Biggr \}^{1-v^2}\,
\label{func2}
\eeq
The asymptotic expression for the wave function $\phi^{a}$ are given as follows \\
 for the leading twist 2 wave functions:
\beq
P_A(x)=V_L(x)=V_T(x)=\phi^{a}(x)=6x_1 x_2\,,
\eeq
for the non-leading twist 3 wave functions:
\beq
 P_P(x)=1,\quad V_{\perp}(x)=\frac{3}{4}[1+(x_1-x_2)^2]\,,
\eeq
\beq
V_A(x)=P_T(x)=6 x_1 x_2\,.\nonumber
\eeq
The expression for light cone wave function (\ref{func1}), (\ref{func2}) is one of the 
possible ways to link different limits: quark-antiquark pair in meson being in the rest $v \to 0$ and 
very light quark $v \to 1$. In the former limit one obviously gets $\sim \delta (x - 1/2)$, the later one 
leads to the function $\sim \phi^{a}$.

In the numerical analysis the following parameters will be used:
\beq
\nonumber
\overline {M}_c = 1.2 \mbox{GeV}, \\ \nonumber
\JP(1S), \ec(1S) ~~ |f_P| \simeq |f_V| \simeq 0.41 \mbox{GeV},  \\ 
\JP(2S), \ec(2S) ~~ |f_P| \simeq |f_V| \simeq 0.28 \mbox{GeV},
\label{const}
\eeq
The values $f_V$ was obtained from decay width $\Gamma (V \to e^+ e^-)$
\beq
\Gamma(V \to e^+ e^-) = \frac {16 \pi \alpha^2} {27} \frac {|f_V|^2} {M_V}.
\eeq
The constants $f_V, f_P$ are considered to be equal: $f_P \simeq f_V $.
For $\alpha_s(\mu)$ one loop result will be used
\beq
\alpha_s(\mu)=\frac {4 \pi} {b_0 \log (\mu^2/ \Lambda^2)},
\eeq
with $\Lambda = 200$MeV. Last parameter needed for numerical analysis is the width of wave function $v^2$.
It will be taken from potential models\cite{Buchmuller:1980su}: 
\beq
\nonumber
\JP, \ec ~~ v^2=0.23 , \\
\JP(2S), \ec(2S) ~~ v^2=0.29.
\eeq
The result of the calculation is presented in Table 1. The second and the third columns 
contain experimental result measured at Babar and Belle experiments.
In the fourth column the results of this section are presented. In order to compare the 
result with NRQCD predictions for the processes under consideration the fifth 
column contains the predictions in the framework of this model. 

From table 1 one sees that the predictions of the cross section of the processes 
$\epem\to\JP\ec, \JP(2S) \ec, \JP \ec(2S), \JP(2S) \ec(2S)$ in the framework of light cone 
is much greater than NRQCD predictions. As was noted above the reason of this 
discrepancy may be attributed to the fact that at leading approximation NRQCD 
does not regard the motion inside final mesons. In paper \cite{Braaten:2002fi} it was noted that 
NRQCD corrections to the amplitude with final $\JP(2S), \ec(2S)$ mesons are 
large(expansion parameter in this case is about $v^2 \sim 0.7$). So the 
application of NRQCD to this processes is unreliable. Moreover strong 
dependence of the amplitude from the propagators of the intermediate particles
mentioned above does not improve NRQCD also. As the result NRQCD prediction is 
in poor agreement with the experiment data. In contrast to NRQCD the leading order light cone predictions
is in better agreement with data, from what one may suppose that light cone expansion is more reliable for 
$\epem\to V P$ at energy $\sqrt s =10.6$GeV. The problem with light cone expansion 
is connected with poor knowledge of light cone wave function (\ref{func1}),(\ref{func2})
especially in the case of $2S$ meson. And in order to get better understanding of the processes under consideration 
in addition to QCD corrections and next to leading term in $1/s$ expansion one 
should obtain better knowledge of wave functions (\ref{func1}),(\ref{func2}).

There are two contributions to the  formfactor $F_{vp}$ separated in formula (\ref{res}).
The first contribution originates from the wave function of mesons at the origin and it is 
proportional to $\sim f_V f_P$.  The second contribution regards internal motion 
of quark-antiquark pair inside mesons and it is proportional to $I_0$. As was noted 
above leading   NRQCD approximation does not take into the account contribution of 
the second type. Moving form the lower $c \bar c$ states to the upper ones 
we diminish the value of the constant $f_V, f_P$. So in the framework of NRQCD 
the cross sections for the production of upper $c \bar c$ mesons is less than that 
for the lower $c \bar c$ states. This effect is well seen in table 1. 
The second contribution $\sim I_0$ where the internal motion is taken into account 
compensate first effect since upper lying resonances are broader. 
To the first sight one may conclude that the cross section of the processes
$\epem\to\JP(2S) \ec$ and $\epem\to \JP \ec(2S)$ can not differ significantly. 
But our calculations show that the cross section $\epem\to \JP \ec(2S)$ is 
almost two times larger than that for $\epem\to \JP(2S) \ec$. The reason of such large 
discrepancy consists in the fact that some terms in formula \ref{res} is multiplied by 
the factor $M_P/M_V$. This factor enhances or diminishes different terms in (\ref{res})
what results in the enhancement of the $\sigma (\epem\to \JP \ec(2S) )$ in comparison with $ \sigma ( \epem\to \JP(2S) \ec )$.
This fact is a peculiarity of light cone prediction.

\section{The processes $\epem\to\JP(1S) \chi_{c0}, \JP(2S) \chi_{c0}$. }

Leading  asymptotic behavior of matrix element $\left<V(p_1, \lambda), S(p_2)| J_{\mu} | 0\right>$ may be derived from formula (\ref {as}).
Here we have $M(p_1, \lambda_1)=V(p_1, \lambda), M(p_2, \lambda_2)=S(p_2)$. Obviously the helicity $\lambda_2$ equals zero. As 
to the vector meson $V$ the leading contribution is given by the helicity $\lambda_1=0$. So the asymptotic behavior of the amplitude is
\beq
\left<V(p_1, \lambda), S(p_2)| J_{\mu} | 0\right> & \sim & \frac 1 {\sqrt s}.
\eeq
Then the asymptotic behavior of the cross section $\sigma (\epem\to VS)$ is $\sim 1/s^3$. Unfortunately leading 
in $1/s$ expansion contribution is much less than NLO. To prove this NRQCD result for the 
cross section of the processes under consideration obtained in paper \cite{Braaten:2002fi}  
will be used. Let's consider the process $\epem\to\JP(1S) \chi_{c0}$.
Cross section of this processes can be represented in the form
\beq
\sigma & = & \frac {\pi^3} {3^5 s} \alpha^2 \alpha_s^2 q_c^2 F_0 r^2 \sqrt{1-r^2} \frac {f_V^2 f_S^2} {m_c^4}.
\eeq
where $r^2 = 16 m_c^2 /s$ and $F_0 = 2(18 r^2 - 7 r^4)^2+ r^2(4 + 10 r^2 - 3 r^4)^2$. Let us substitute 
$s \to 10.6^2 t$ and expand the above formula in $1/t$ series ($m_c=1.4$GeV). We get
\beq
\sigma & = & \frac {0.15} {t^3} + \frac {1.84} {t^4} + O(1/t^5).  
\eeq
Thus one sees that in the framework of NRQCD NLO correction   at energy $\sqrt s = 10.6$ GeV  is an order of magnitude larger
than that leading one. In light cone NRQCD result is multiplied by the factor that accounts internal motion. Thus 
if one supposes that these factors are of the same order of magnitude for LO and NLO contributions one may conclude that 
NLO contribution in the framework of light cone is much larger than LO. The same is true for the process 
$\epem\to \JP(2S) \chi_{c0}$. Leading order result can be found in paper \cite{Chernyak:1983ej}. NLO contribution will be considered in our forthcoming publication.

\section{Discussion.}

In this paper we have reanalyzed the production of excited charmonium mesons pair in  $\epem$ annihilation  
(i.e. the reactions $\epem\to \psi(1S)\eta_c(1S), \psi(2S)\eta_c(1S), \psi(1S)\eta_c(2S),\psi(2S)\eta_c(2S)$ and $\psi\chi_{c0}$) 
in the framework of the light cone. It is shown, that the internal motion of quarks inside charmonium mesons leads to 
substantial increase of the cross sections of these processes and reasonable agreement between theoretical predictions and 
available experimental data can be reached. 
It is also shown, that at energy $\sqrt s =10.6$GeV NLO contribution to the cross sections $\epem\to\psi(1S)\chi_{c0}, \psi(2S)\chi_{c0}$ are about an order of magnitude higher, 
than the LO contribution. So in order to achieve agreement between theoretical predictions of the cross sections of the 
processes $\epem\to\psi(1S)\chi_{c0}, \psi(2S)\chi_{c0}$ and experimental data  one should calculate NLO contribution to the cross sections.

The authors thank V.L. Chernyak for useful discussions. This work was partially
supported by Russian Foundation of Basic Research under grant 04-02-17530, Russian Education
Ministry grant E02-31-96, CRDF grant MO-011-0, Scientific School grant SS-1303.2003.2. One of
the authors (V.B.) was also supported by Dynasty foundation.

\newpage

\begin{table}
$$\begin{array}{|c|c|c|c|c|}
\hline
   H_1 H_2 & \sigma_{BaBar} \times Br_{H_2 \to charged >2}(fb)  \mbox{\cite{Abe:2004ww}} 
 & \sigma_{Belle}\times Br_{H_2 \to charged >2}(fb)  \mbox{\cite{Aubert:2005tj}}
 & \sigma_{LO} (fb)  
 & \sigma_{NRQCD} (fb) \mbox{\cite{Braaten:2002fi}}\\
\hline
\JP(1S) \ec(1S)& 17.6 \pm 2.8^{+1.5}_{-2.1} & 25.6 \pm 2.8 \pm 3.4 & 26.7  &2.31\\
\hline
\JP(2S) \ec(1S)&  - & 16.3 \pm 4.6 \pm 3.9 & 16.3 &0.96\\
\hline
 \JP(1S) \ec(2S)& 16.4 \pm 3.7^{+2.4}_{-3.0} & 16.5 \pm 3.0 \pm 2.4 & 26.6 &0.96\\
\hline
 \JP(2S) \ec(2S)& - & 16.0 \pm 5.1 \pm 3.8 & 14.5 &0.40\\
\hline
\end{array}$$
\end{table}

\begin{figure}[ph]
\begin{picture}(150, 200)
\put(-100,-0){\epsfxsize=9cm \epsfbox{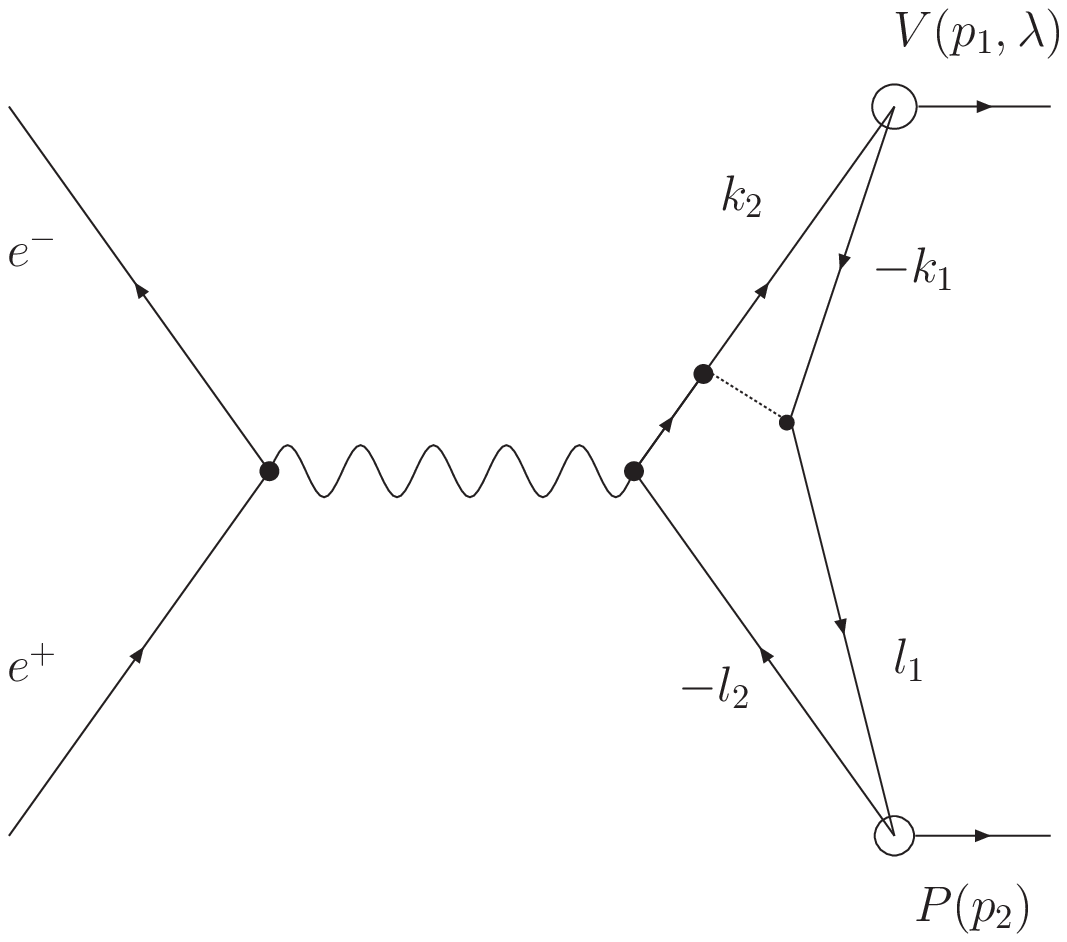}}
\put(50,0){\bf{Fig. 1a}}
\put(-100,-250){\epsfxsize=9cm \epsfbox{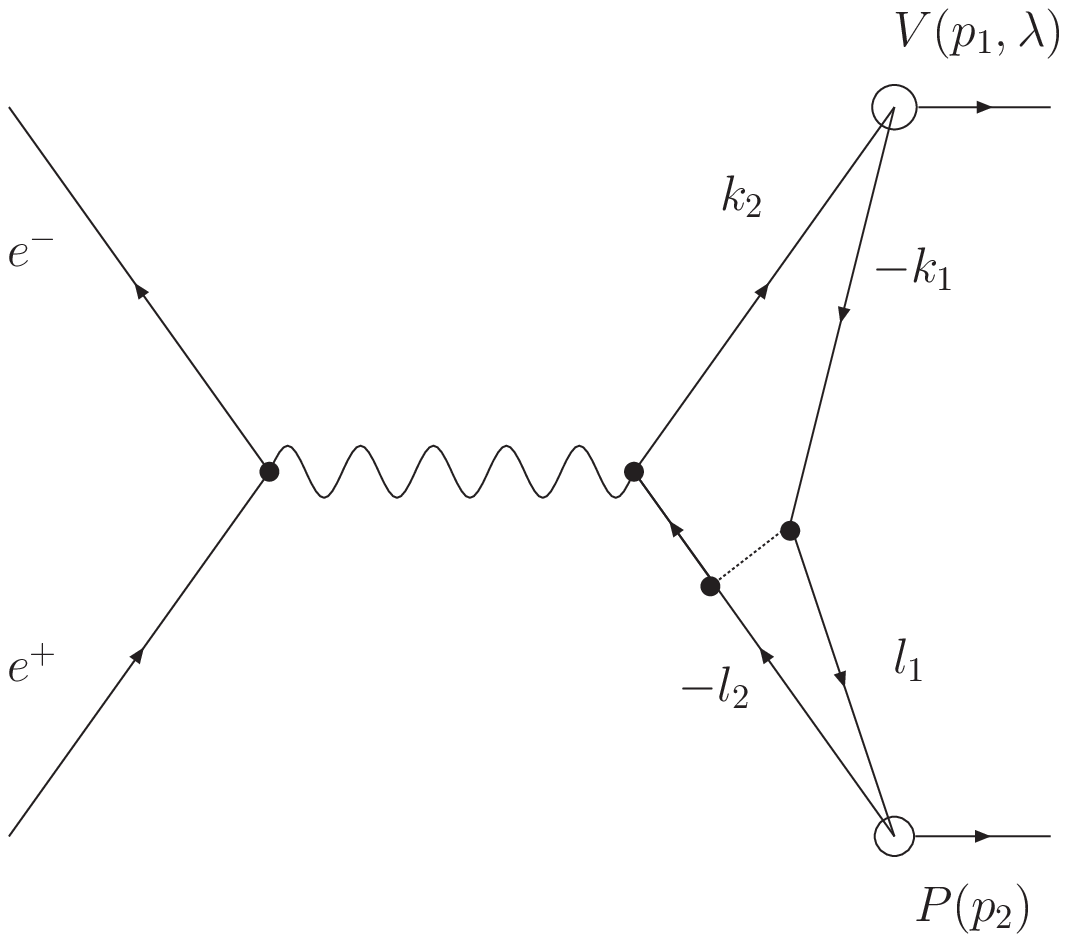}}
\put(50,-250){\bf{Fig. 1b}}
\end{picture}
\end{figure}


\begin{thebibliography}
\expandafter\ifx\csname natexlab\endcsname\relax\def\natexlab#1{#1}\fi
\expandafter\ifx\csname bibnamefont\endcsname\relax
  \def\bibnamefont#1{#1}\fi
\expandafter\ifx\csname bibfnamefont\endcsname\relax
  \def\bibfnamefont#1{#1}\fi
\expandafter\ifx\csname citenamefont\endcsname\relax
  \def\citenamefont#1{#1}\fi
\expandafter\ifx\csname url\endcsname\relax
  \def\url#1{\texttt{#1}}\fi
\expandafter\ifx\csname urlprefix\endcsname\relax\def\urlprefix{URL }\fi
\providecommand{\bibinfo}[2]{#2}
\providecommand{\eprint}[2][]{\url{#2}}

\bibitem[{\citenamefont{Abe et~al.}(2002)}]{Abe:2002rb}
\bibinfo{author}{\bibfnamefont{K.}~\bibnamefont{Abe}} \bibnamefont{et~al.}
  (\bibinfo{collaboration}{Belle}), \bibinfo{journal}{Phys. Rev. Lett.}
  \textbf{\bibinfo{volume}{89}}, \bibinfo{pages}{142001}
  (\bibinfo{year}{2002}), \eprint{hep-ex/0205104}.

\bibitem[{\citenamefont{Braaten and Lee}(2003)}]{Braaten:2002fi}
\bibinfo{author}{\bibfnamefont{E.}~\bibnamefont{Braaten}} \bibnamefont{and}
  \bibinfo{author}{\bibfnamefont{J.}~\bibnamefont{Lee}},
  \bibinfo{journal}{Phys. Rev.} \textbf{\bibinfo{volume}{D67}},
  \bibinfo{pages}{054007} (\bibinfo{year}{2003}), \eprint{hep-ph/0211085}.

\bibitem{Bodwin:1994jh}
  G.~T.~Bodwin, E.~Braaten and G.~P.~Lepage,
  Phys.\ Rev.\ D {\bf 51} (1995) 1125
  [Erratum-ibid.\ D {\bf 55} (1997) 5853],
  hep-ph/9407339.




\bibitem[{\citenamefont{Bodwin et~al.}(2003{\natexlab{a}})\citenamefont{Bodwin,
  Lee, and Braaten}}]{Bodwin:2002fk}
\bibinfo{author}{\bibfnamefont{G.~T.} \bibnamefont{Bodwin}},
  \bibinfo{author}{\bibfnamefont{J.}~\bibnamefont{Lee}}, \bibnamefont{and}
  \bibinfo{author}{\bibfnamefont{E.}~\bibnamefont{Braaten}},
  \bibinfo{journal}{Phys. Rev. Lett.} \textbf{\bibinfo{volume}{90}},
  \bibinfo{pages}{162001} (\bibinfo{year}{2003}{\natexlab{a}}),
  \eprint{hep-ph/0212181}.

\bibitem[{\citenamefont{Bodwin et~al.}(2003{\natexlab{b}})\citenamefont{Bodwin,
  Lee, and Braaten}}]{Bodwin:2002kk}
\bibinfo{author}{\bibfnamefont{G.~T.} \bibnamefont{Bodwin}},
  \bibinfo{author}{\bibfnamefont{J.}~\bibnamefont{Lee}}, \bibnamefont{and}
  \bibinfo{author}{\bibfnamefont{E.}~\bibnamefont{Braaten}},
  \bibinfo{journal}{Phys. Rev.} \textbf{\bibinfo{volume}{D67}},
  \bibinfo{pages}{054023} (\bibinfo{year}{2003}{\natexlab{b}}),
  \eprint{hep-ph/0212352}.

\bibitem[{\citenamefont{Luchinsky}(2003)}]{Luchinsky:2003yh}
\bibinfo{author}{\bibfnamefont{A.~V.} \bibnamefont{Luchinsky}},
  \bibinfo{journal}{Atom. Nucl.hys. Atom. Nucl.} \textbf{\bibinfo{volume}{67}},
  \bibinfo{pages}{1338} (\bibinfo{year}{2003}), \eprint{hep-ph/0301190}.

\bibitem[{\citenamefont{Abe et~al.}(2003)}]{Abe:2003ja}
\bibinfo{author}{\bibfnamefont{K.}~\bibnamefont{Abe}} \bibnamefont{et~al.}
  (\bibinfo{collaboration}{Belle}) (\bibinfo{year}{2003}),
  \eprint{hep-ex/0306015}.


\bibitem{Lee:2003db}
  J.~Lee,
  J.\ Korean Phys.\ Soc.\  {\bf 45}, S354 (2004),
  hep-ph/0312251.



\bibitem{Brambilla:2004wf}
  N.~Brambilla {\it et al.}  [QWG],
  hep-ph/0412158.

\bibitem{Zhang:2005ch}
  Y.~J.~Zhang, Y.~j.~Gao and K.~T.~Chao,
  hep-ph/0506076.


\bibitem[{\citenamefont{Ma and Si}(2004)}]{Ma:2004qf}
\bibinfo{author}{\bibfnamefont{J.~P.} \bibnamefont{Ma}} \bibnamefont{and}
  \bibinfo{author}{\bibfnamefont{Z.~G.} \bibnamefont{Si}},
  \bibinfo{journal}{Phys. Rev.} \textbf{\bibinfo{volume}{D70}},
  \bibinfo{pages}{074007} (\bibinfo{year}{2004}), \eprint{hep-ph/0405111}.

\bibitem[{\citenamefont{Bondar and Chernyak}(2004)}]{Bondar:2004sv}
\bibinfo{author}{\bibfnamefont{A.~E.} \bibnamefont{Bondar}} \bibnamefont{and}
  \bibinfo{author}{\bibfnamefont{V.~L.} \bibnamefont{Chernyak}}
  (\bibinfo{year}{2004}), \eprint{hep-ph/0412335}.
  
\bibitem{Braguta:2005gw}
  V.~V.~Braguta, A.~K.~Likhoded and A.~V.~Luchinsky,
  hep-ph/0506009.



\bibitem[{\citenamefont{Buchmuller and Tye}(1981)}]{Buchmuller:1980su}
\bibinfo{author}{\bibfnamefont{W.}~\bibnamefont{Buchmuller}} \bibnamefont{and}
  \bibinfo{author}{\bibfnamefont{S.~H.~H.} \bibnamefont{Tye}},
  \bibinfo{journal}{Phys. Rev.} \textbf{\bibinfo{volume}{D24}},
  \bibinfo{pages}{132} (\bibinfo{year}{1981}).




\bibitem{Abe:2004ww}
  K.~Abe {\it et al.}  [Belle Collaboration],
  Phys.\ Rev.\ D {\bf 70}, 071102 (2004),
  hep-ex/0407009.





\bibitem{Aubert:2005tj}
  B.~Aubert  [BABAR Collaboration],
  hep-ex/0506062.



\bibitem{Chernyak:1977as}
  V.~L.~Chernyak and A.~R.~Zhitnitsky,
  JETP Lett.\  {\bf 25}, 510 (1977)
  [Pisma Zh.\ Eksp.\ Teor.\ Fiz.\  {\bf 25}, 544 (1977)].

\bibitem[{\citenamefont{Chernyak and Zhitnitsky}(1984)}]{Chernyak:1983ej}
\bibinfo{author}{\bibfnamefont{V.~L.} \bibnamefont{Chernyak}} \bibnamefont{and}
  \bibinfo{author}{\bibfnamefont{A.~R.} \bibnamefont{Zhitnitsky}},
  \bibinfo{journal}{Phys. Rept.} \textbf{\bibinfo{volume}{112}},
  \bibinfo{pages}{173} (\bibinfo{year}{1984}).


\end{thebibliography}
\end{document}